\documentclass[journal]{IEEEtran}

\usepackage{xcolor,soul,framed} 

\colorlet{shadecolor}{yellow}
\usepackage[pdftex]{graphicx}
\graphicspath{{../pdf/}{../jpeg/}}
\DeclareGraphicsExtensions{.pdf,.jpeg,.png}

\usepackage[cmex10]{amsmath}

\usepackage{array}
\usepackage{mdwmath}
\usepackage{mdwtab}
\usepackage{eqparbox}
\usepackage{url}
\usepackage{multicol}
\usepackage{float}
\usepackage{booktabs}
\usepackage{rotating}
\usepackage[style=ieee]{biblatex}
\begin{document}

\bstctlcite{IEEEexample:BSTcontrol}
    \title{Exploring the Efficacy of Large Language Models (GPT-4) in Binary Reverse Engineering}

\author{
\begin{tabular}{@{}c@{\hspace{1em}}c@{}}
\centering
Saman Pordanesh & Dr. Benjamin Tan \\
saman.pordanesh@ucalgary.ca & benjamin.tan1@ucalgary.ca
\end{tabular}
}

\markboth{UNIVERSITY OF CALGARY, SCHULICH SCHOOL OF ENGINEERING, UNDERGRADUATE RESEARCH THESIS, FALL 2023
}{S. Pordanesh}

\maketitle

\begin{abstract}

This study investigates the capabilities of Large Language Models (LLMs), specifically GPT-4 \cite{gpt4}, in the context of Binary Reverse Engineering (RE). Employing a structured experimental approach, we analyzed the LLM's performance in interpreting and explaining human-written and decompiled codes. The research encompassed two phases: the first on basic code interpretation and the second on more complex malware analysis. Key findings indicate LLMs' proficiency in general code understanding, with varying effectiveness in detailed technical and security analyses. The study underscores the potential and current limitations of LLMs in reverse engineering, revealing crucial insights for future applications and improvements. Also, we examined our experimental methodologies, such as methods of evaluation and data constraints, which provided us with a technical vision for any future research activity in this field.

\end{abstract}


\IEEEpeerreviewmaketitle
\section{Introduction}

\IEEEPARstart{T}{he} advent of Large Language Models (LLMs), a groundbreaking AI technology employing deep learning, has revolutionized the way we interact with machines. These advanced models can understand human language and generate relevant, context-specific content for various Natural Language Processing (NLP) \cite{nlp} tasks. This advancement has sparked considerable interest across diverse professional fields as people seek to leverage these tools to enhance the efficiency and effectiveness of their work.

In computer science, particularly in reverse engineering, challenges in code comprehension and pattern recognition are common. LLMs, however, are showing promising capabilities in various computer science-related tasks, including interpreting and explaining code using these sophisticated deep learning models \cite{sarsa2022automatic}.

Reverse engineering of binary files, a critical aspect of computer science, involves deciphering the underlying code and design logic of binary files. This process is naturally complex and time-consuming \cite{anderson2014automating}, as it requires retracing the steps of code compilation. The potential application of LLMs in Binary Reverse Engineering sparked our curiosity, raising questions about their performance, current capabilities, and areas for further improvement in this challenging field. A key area of interest is the application of LLMs in understanding decompiled binary code, a task markedly different from interpreting standard, human-written code.

To address these questions, our research seeks to establish a meaningful connection between reverse engineering tasks and LLMs' capabilities. We aim to explore these two domains' intersections, enabling us to conduct more in-depth investigations and address more specific, real-world challenges.

This report presents the initial stage of our research journey. We delve into various infrastructures, application perspectives, and the scope of coverage in this field. Additionally, we survey the technical aspects of our experimental methods, including evaluation, implementation, and the constraints of the tools. Our goal is to provide a comprehensive understanding of the role and potential of LLMs in building up Binary Reverse Engineering processes.

\subsection{Overview of the Literature}
We did a detailed literature survey of at least ten related papers and provided a strong background of what we had done in the experimental phase. The landscape of AI-assisted reverse engineering and code analysis has seen significant advancements, as evidenced by recent research. Studies like those by Pearce et al. \cite{pearce2022pop} and Wong et al. \cite{wong2023refining} have demonstrated the potential of large language models like OpenAI's Codex \cite{openaicodex} to enhance the understanding and recompilability of decompiled code. These advancements suggest significant matches between AI tools and traditional reverse engineering techniques, suggesting a good option for addressing complex software analysis and cybersecurity challenges. Furthermore, the efforts by Davis et al. \cite{davis2023feasibility}, and Al-Kaswan et al. \cite{al2023extending} in binary code summarization underscore the burgeoning role of AI in generating concise, understandable summaries of complex binary code. This is crucial in a landscape where the volume and complexity of malware and software rapidly expand.

Moreover, the application of AI in educational contexts, as explored by Sarsa et al. \cite{sarsa2022automatic}, and in malware analysis, as developed by Novkovicé et al. \cite{novkovic2016can}, and Hajipour et al. \cite{hajipour2021ireen}, highlight the versatility of AI tools. Creating specialized datasets and models, such as BinT5 \cite{bint52023} and MLARES, demonstrate the expanding capabilities of AI in providing insightful analysis and support in diverse areas of programming and cybersecurity. This body of work collectively indicates a significant shift towards integrating advanced AI techniques in software engineering, offering innovative solutions to longstanding challenges in code analysis, malware classification, and programming education.

\subsection{Our Research Direction}
Based on our literature survey, which we talked about briefly in the previous section, experiments were discovering LLM's capability in some areas like code explanations \cite{sarsa2022automatic, wong2023refining, al2023extending}, recompiling a decompiled code \cite{wong2023refining, al2023extending}, and training LLMs for security tasks \cite{nadeem2023sok, novkovic2016can, davis2023feasibility}.

Generally, these publications focus on machine learning and neural network capability to ease reverse engineering tasks like malware analysis and malicious code or file recognition. We wrapped our thoughts around these different fields and decided to be more focused on code explanation, which keeps this research close to the code understanding aspect of the LLMs. The reason is that firstly, we found these areas have more capacity for improvement and make some RE jobs easier. Also, some RE tasks like malware detection are already being researched and developed by well-known cybersecurity companies for commercial purposes. However, without going deep into malware detection, papers around this field provided us with helpful materials on research jobs like data collection, evaluation methods, and analysis approaches.

\section{Setups \& Tools}
This section outlines our study's specific tools and technologies, detailing the range of tools and technological frameworks utilized. It also illuminates our data collection strategies throughout the experiment.

\subsection{Model Selection - GPT4}
Central to this experiment is exploring Large Language Model (LLM) performance. With various models available, including OpenAI's Codex \cite{openaicodex}, GPT-3.5 \cite{gpt-3.5}, GPT-4 \cite{gpt4}, and Facebook's LlaMa \cite{llama}, our choice fell on GPT-4. Selected based on benchmarks Trustbit LLM Benchmark \cite{holistic, peng2023instruction, gemini}, GPT-4 stands out as one of the most advanced LLMs in NLP, positioning us to evaluate the forefront of this technology and later form robust recommendations for improvement.

\subsection{Tools}
\subsubsection{Python Scripting and Data Storage}
The experiment interfaces with GPT-4 via its API using a unique API key. All operations, including data transformation, prompt creation, communication with the model, and result collection, are conducted through Python scripting. Data, encompassing examination details, outcomes, and evaluation results, are stored in \textbf{.csv} format for simplicity. While adequate for this study's scale, more extensive future studies may necessitate more advanced data storage solutions. Our GitHub repository \cite{pordanesh2023} consolidates all scripts and data, ensuring effective version control.

\subsubsection{Decompiling Tools - Ghidra and RetDec}
We utilized two distinct open-source decompiling methods: Ghidra \cite{ghidra} and RetDec \cite{retdec}. Ghidra, an NSA-developed tool, offers decompiling, disassembling, and code language detection capabilities. RetDec, on the other hand, supports various programming languages and processor architectures. Both tools were crucial in our experiment for different scenarios.

\subsection{Data}
We discuss our data collection strategies and methods here, though subsequent sections will provide more detailed descriptions of the data specific to each phase.

\subsubsection{Dataset 1 - Simple C Programming Problems}

This dataset comprises 70 simple C programming problems sourced from Programiz \cite{programiz}. These examples range from 20 to 60 lines and are stored in a \textbf{.csv} file with columns for problem name, explanatory notes, original code, and a comment-free version of the code. This dataset primarily supports the first phase of our experiment, with additional data points added for each scenario as needed.

\subsubsection{Dataset 2 - Malware Source Codes in C}

The second dataset features 15 malware source codes in C, all obtained from open-source GitHub \cite{github} repositories, complete with comments and documentation. These codes underwent a manual compile-decompile process, yielding two versions of decompiled code using Ghidra \cite{ghidra} and RetDec \cite{retdec} decompilers. Stored in a \textbf{.csv} file, the data includes columns for the source, platform applicability, original code, and both versions of decompiled code. This dataset was integral to the latter phases of our study.

\section{Experiment Design}
In this section, we detail the various phases and scenarios that represent the experimental journey. Each phase is carefully designed to explore distinct aspects of Large Language Models (LLMs) in Binary Reverse Engineering. We will explain the objectives of each phase, the methodologies employed, and the intended outcomes of these investigative exercises.

This experiment is structured into two distinct phases, each containing various scenarios tailored to diverse evaluation perspectives. This diversity ensures a comprehensive exploration of the subject matter, focusing on specific objectives and methodologies in each phase.

More details like scripts, data, prompts, and rubrics can be found on the project GitHub repository \cite{pordanesh2023} and the documentation there.

\subsection{Phase 1: Basic Code Interpretation}

\begin{figure*}[t]
    \centering
    \includegraphics[width=\textwidth]{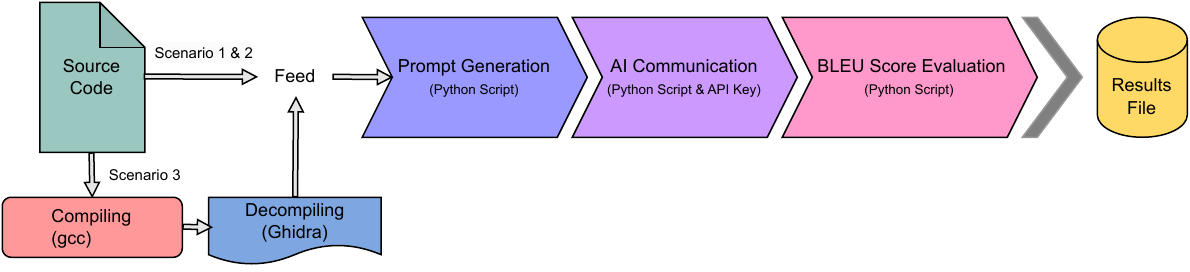}
    \caption{Workflow Diagram of Phase 1 - Basic Code Interpretation and Analysis in Large Language Model Experimentation}
    \label{fig:fig1-ph1}
\end{figure*}

Phase 1 lays the foundational groundwork for our experiment, focusing on establishing the infrastructure, refining data-gathering methodologies, and assessing the basic capabilities of GPT-4 in explaining code. This phase is divided into three distinct scenarios designed to evaluate the model's performance. Please see Figure~\ref{fig:fig1-ph1} for a visualized workflow in this phase. 

Data Source: We selected 70 examples of C programming exercises from Programiz \cite{programiz}, ranging from essential "Hello World" programs to intermediate-level geomatics calculations. These examples, typically used in C programming tutorials, perfectly align with our objective of simplicity for this initial phase.

Objective: The primary goal in this phase is to observe and analyze how LLMs, GPT-4 \cite{gpt4}, respond to straightforward requests for code explanations. We aim to gain insights into the model's interpretative skills and foundational understanding of programming concepts when it faces human written code and something more ambiguous like uncommented and decompiled code. This phase serves as a stepping stone for more complex analyses in the next phase.

\subsubsection{Scenario 1: Original Code Explanation}
In Scenario 1, our approach involved feeding the Large Language Model (LLM) the original code from tutorials, complete with comments and in-code documentation. The prompt was carefully tailored to include the application name and programming language, guiding the model to provide concise, precise, and descriptive explanations. The objective of this scenario was to observe the LLM's performance in explaining simple code that includes all the typical aspects of human-written code.

\subsubsection{Scenario 2: Stripped Code Explanation}
Moving to Scenario 2, the approach differed slightly. All comments and documentation were removed from the code while retaining the original function and variable names and their structural integrity. The prompt was, on purpose, made more ambiguous by omitting the function application name and the reference to "C programming." The aim was to assess the LLM's capability to interpret code with reduced contextual hints, thereby evaluating its adaptability and understanding in somewhat ambiguous settings.

\subsubsection{Scenario 3: Decompiled Code Explanation}
In Scenario 3, we took a different route. All code samples were compiled and then decompiled using Python automation along with Ghidra \cite{ghidra} as the decompilation tool. The decompiled functions then served as the basis for generating prompts for each data entry, maintaining a descriptiveness similar to that of Scenario 2. The primary focus here was to test the LLM's ability to explain decompiled code, which lacks meaningful names, comments, documentation, or structure, posing a more significant challenge in code explanation.

\subsubsection{Evaluation: Unified Approach Across Scenarios}
Finally, a consistent evaluation methodology was applied across all scenarios. Utilizing Python scripting for automation, we compared the explanations generated by the LLM against the original explanations sourced from the tutorials, employing the BLEU score \cite{bleu} for this purpose. The BLEU score, commonly used in translation quality assessment, enabled us to quantify the similarity between the LLM's output and the reference explanations. This unified approach not only streamlined the comparison process but also critically evaluated the effectiveness of BLEU scoring in the context of this experimental study.

\subsection{Phase 2: Malware Analysis}
\begin{figure*}[t] 
    \centering
    \includegraphics[width=\textwidth]{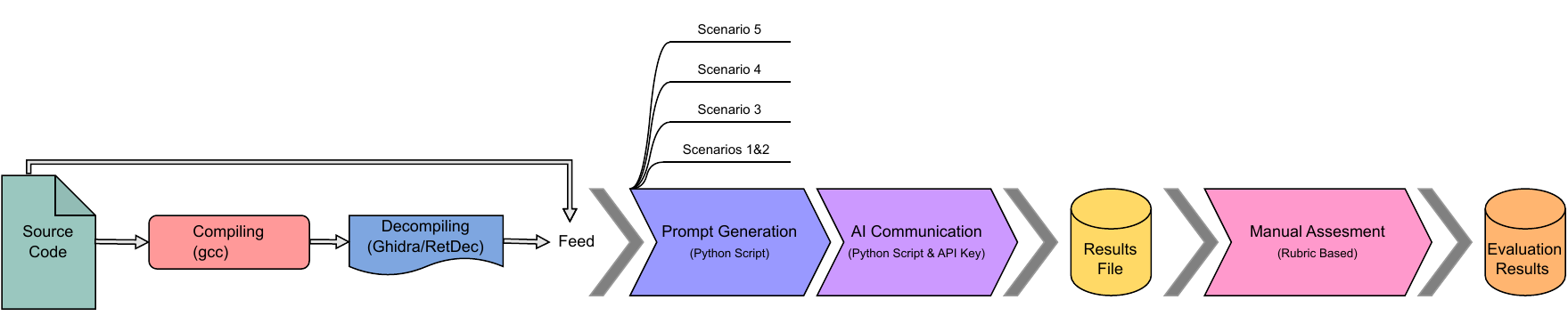}
    \caption{ Workflow Diagram of Phase 2 - Advanced Analysis of Malware-Reversed Engineered Applications Using Large Language Models}
    \label{fig:fig1-ph2}
\end{figure*}

In Phase 2, our investigation explores the LLM's proficiency in explicating real-world, malware-reversed engineered applications, mainly written in C. This phase was carefully structured with specific scenarios to comprehensively evaluate the LLM's capabilities in code explanation under varying conditions.

Dataset Composition: We collected a set of 15 malware source codes from publicly accessible GitHub \cite{github} repositories. The selection criteria were centered around the extent of in-code comments and documentation, ensuring a rich dataset for analysis. This selection aims to mirror real-world complexities experienced in reverse engineering scenarios. Please see Figure \ref{fig:fig1-ph2} for a visualized workflow in
this phase

Decompilation Process: All sourced codes were compiled in a secure, isolated Linux environment and subsequently decompiled using RetDec \cite{retdec}. The choice of RetDec over Ghidra \cite{ghidra} for this phase was driven by RetDec's user-friendliness and ability to generate cleaner decompiled code, making future manual assessments easier.

Scenario Design: To thoroughly assess the LLM's performance, we developed various scenarios differing in dataset size, quality, and complexity. These scenarios were strategically designed to understand the different faces of the LLM's code explanation abilities. The following sections provide more information about the objectives and methodologies for each scenario.

\subsubsection{Scenarios 1\&2: Variable and Function Name Detection}

In these initial scenarios, we evaluated the LLM's proficiency in renaming decompiled functions and variables to more human-readable, meaningful, and contextually relevant names. For each decompiled malware application, we crafted a prompt that included the code and requested the LLM to suggest new names for functions and variables, ensuring they are meaningful and align with the overall application context. The results were compiled in JSON format, documenting original and suggested names.

The responses were assessed using a manual scoring method based on a rubric that considered the meaningfulness and effectiveness of the newly suggested names.

\subsubsection{Scenario 3: Enhanced Clarity and Structure}

In Scenario 3, our approach involved presenting the LLM with eight binary (Yes/No) questions applied to each decompiled code sample. This experiment has replicated the original code, which has a more precise and descriptive code structure, including in-code comments and original function/variable names. The purpose was to assess the LLM's ability to provide binary responses that accurately reflect the code's attributes and functionalities.

The questions were carefully chosen to cover a range of analyses, including the program's primary functions, network interactions, scalability, and potential security vulnerabilities. Specific areas of focus were the detection of suspicious API usage, unauthorized network activities, and unnecessary encryption practices. Furthermore, the questions delved into whether the program showed any signs of being stealthy or having persistence mechanisms, such as running in the background or auto-starting.

The selection of these particular questions was grounded in the fundamental inquiries a reverse engineer would challenge during the initial stages of code analysis. This choice aims to test the LLM's proficiency in basic reverse engineering tasks.

To evaluate the LLM's performance, we compared its responses from the decompiled code against those obtained from the original code. This comparison operated assuming that the LLM could accurately interpret and respond to the questions when passing the original code. This assumption establishes a benchmark for assessing the LLM's effectiveness in understanding and analyzing the more challenging decompiled code form.

\subsubsection{Scenario 4: Enhanced Analysis Approach}

Scenario 4 focuses on the LLM's ability to analyze decompiled code, requiring it to generate a short and general analysis from three angles: Functionality Overview, Key Observations, and Security Analysis. The LLM's task was to produce an analysis within a 200-word limit, mirroring the approach a reverse engineer might take in quickly assessing a decompiled code's features.

The selection of these three perspectives was intentional, aimed at examining the depth and accuracy of the LLM's analysis in critical areas of reverse engineering. To evaluate the LLM's performance, we compared its analyses of the original and decompiled codes, using the original code as a benchmark for accuracy.

\subsubsection{Scenario 5: Comprehensive Code Analysis via Questionnaire}

Scenario 5 builds on the previous scenarios to delve deeper into code analysis and explanation tasks. This scenario introduces a seven-question survey designed to examine the structure and functionality of code. The questions aim to uncover the primary function of the code, the roles of key functions and variables, error handling mechanisms, execution flow, dependencies on external libraries, and potential security issues. This comprehensive set of questions mirrors the analytical process a reverse engineer undertakes when dissecting a decompiled code.

The rationale behind these questions is to capture the core of reverse engineering analysis, focusing on crucial aspects that determine the code's functionality and potential security vulnerabilities. The answers to these questions, obtained from the original and decompiled code, provide a basis for comparison. By assuming the original code responses are accurate, we can effectively measure the LLM's proficiency in understanding and explaining complex code structures and behaviors, an essential skill in reverse engineering.

\subsubsection{Evaluation}

In the evaluation phase of these experiments, we employed a careful manual process with a set of detailed rubrics tailored to each scenario. These rubrics serve as a scoring guide, allowing us to objectively assess the LLM's responses against a set standard – the original code's explanations. Each scenario's unique nature involves distinct scoring methodologies and ranges, ensuring an appropriate and fair evaluation of the LLM's performance in diverse contexts. The results of this evaluation, along with the rubrics and prompts used, will be elaborated in the Results section of the report. For a deeper dive into our methods, readers can refer to the project documentation available on our GitHub repository \cite{pordanesh2023}.

\section{Results and Analysis}
In this section, we delve into the findings from our evaluations across different scenarios, dissecting the outcomes based on the respective methodologies and rubrics employed. Each subsection will interpret the numerical results and explore their implications in the context of our experimental framework.

\subsection{Phase 1}

\begin{figure}[h]
  \centering
  \includegraphics[width=1\columnwidth]{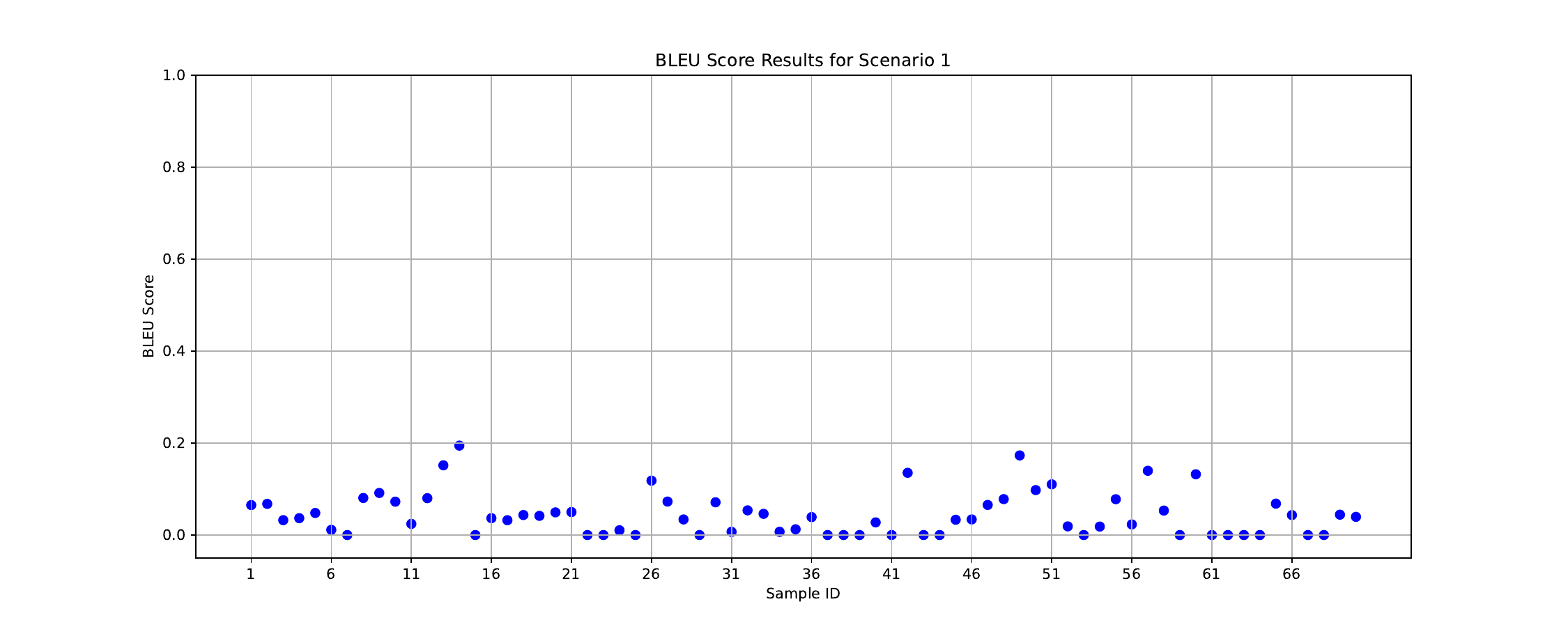}
  \caption{BLEU score results for Scenario 1.}
  \label{fig:scenario1}
\end{figure}

\begin{figure}[h]
  \centering
  \includegraphics[width=1\columnwidth]{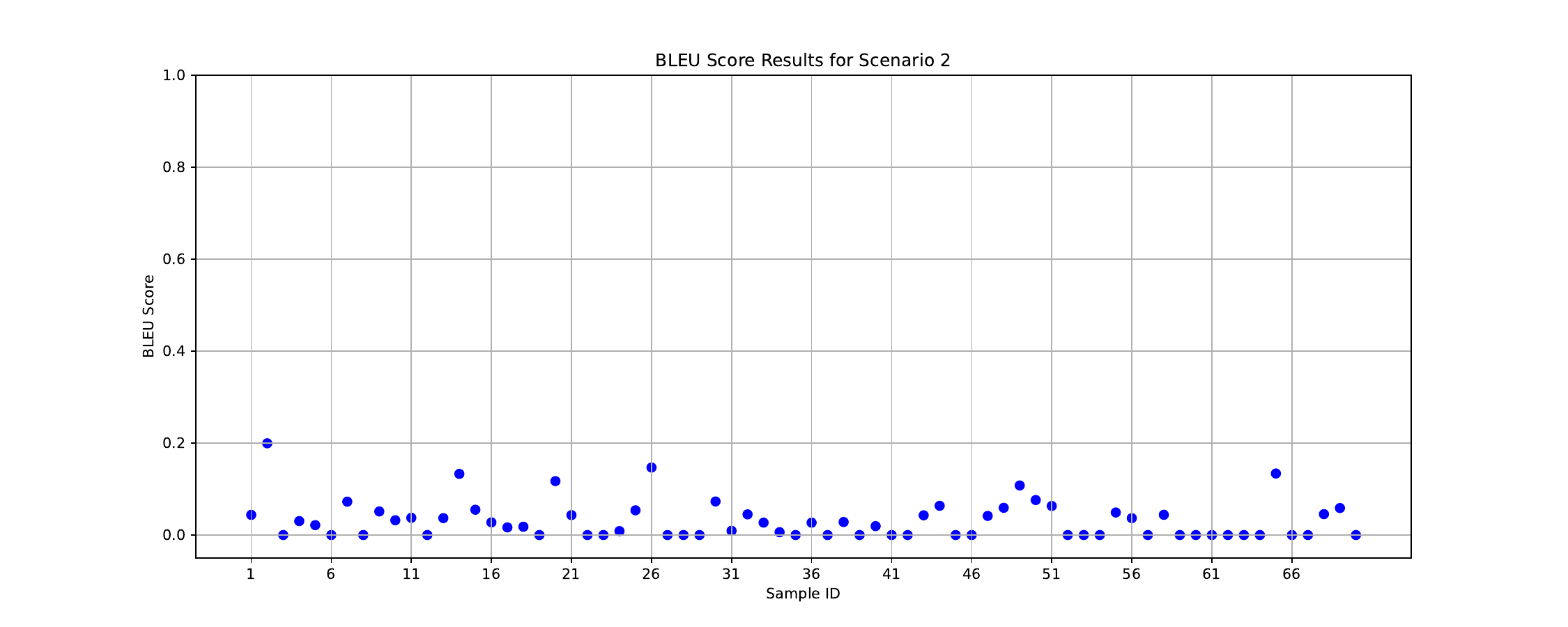}
  \caption{BLEU score results for Scenario 2.}
  \label{fig:scenario2}
\end{figure}

\begin{figure}[h]
  \centering
  \includegraphics[width=1\columnwidth]{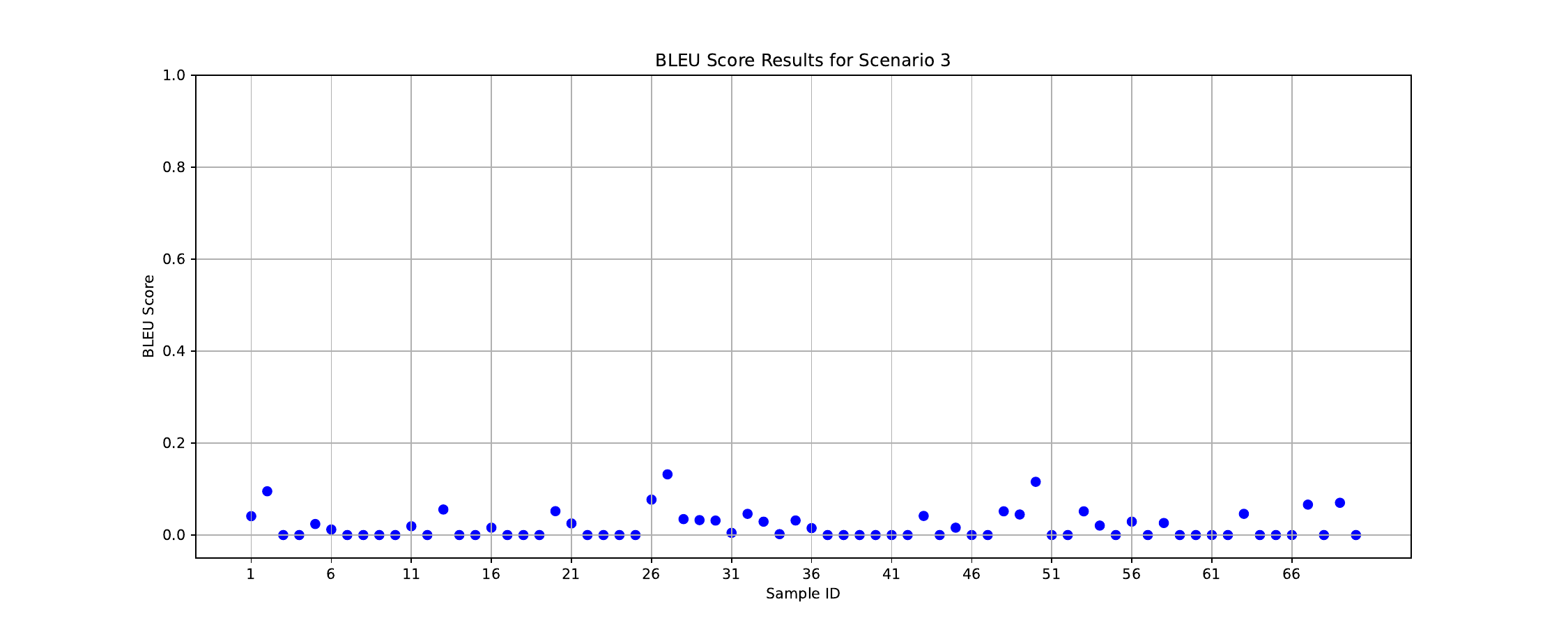}
  \caption{BLEU score results for Scenario 3.}
  \label{fig:scenario3}
\end{figure}

The analysis of BLEU score results (Figures \ref{fig:scenario1}-\ref{fig:scenario3}), computed for each of the 70 code samples across three distinct scenarios, the scores that were revealed were significantly lower than the threshold considered adequate for making meaningful comparisons. According to BLEU score guidelines \cite{bleu}, a score more excellent 0.2 is indicative of a reliable comparison, yet our findings fall short of this benchmark. The accompanying scatter diagrams (Figures \ref{fig:scenario1}-\ref{fig:scenario3}) demonstrate a majority of scores below 0.2, with only a handful approaching this value. This pattern underscores the limitations of relying on automated evaluation methods in their current state for assessing code explanations.

Consequently, we recognize the necessity of abstaining from automated evaluation methods for the time being. While helpful in saving time, these methods require extensive exploration to be effective in the context of code explanation evaluation. Hence, our research path shifts towards Phase 2, prioritizing manual evaluation techniques for a more meaningful and in-depth analysis. This change in strategy indicates a temporary break in analyzing the results of Phase 1, redirecting our focus towards more hands-on evaluation methods in the following phases.

\subsection{Phase 2}
This section will discuss our findings from the manual evaluation of Phase 2, which covered all five scenarios. Each scenario had its unique rubric and scoring system tailored to its complexity and specific aspects.

\subsubsection{Scenario 1\&2}
In these scenarios, we focused on how well GPT-4 \cite{gpt4} could rename functions and variables in 15 decompiled malware codes. We evaluated the effectiveness and meaningfulness of each renamed entity. The meaningfulness score reflected how understandable the new names were for users, while the effectiveness score measured their relevance to other parts of the code.

\begin{table}[H]
\centering
\caption{\textbf{42 Functions} Renaming Task Average Evaluated Scores}
\label{table:functions}
\begin{tabular}{lc}
\toprule
\textbf{Aspect} & \textbf{Average Score (out of 3)} \\
\midrule
Meaningfulness & 2.83 \\
Effectiveness & 2.29 \\
\bottomrule
\end{tabular}
\end{table}

\begin{table}[H]
\centering
\caption{\textbf{69 Variables} Renaming Task Average Evaluated Scores}
\label{table:variables}
\begin{tabular}{lc}
\toprule
\textbf{Aspect} & \textbf{Average Score (out of 3)} \\
\midrule
Meaningfulness & 2.49 \\
Effectiveness & 2.09 \\
\bottomrule
\end{tabular}
\end{table}

In these scenarios, we focused on how well GPT-4 could rename functions (Table \ref{table:functions}) and variables (Table \ref{table:variables}) in 15 decompiled malware codes. We evaluated the effectiveness and meaningfulness of each renamed entity. The meaningfulness score reflected how understandable the new names were for users, while the effectiveness score measured their relevance to other parts of the code.

The higher scores for function names compared to variable names suggest that GPT-4 might be more aware of the code's general context, aiding it in choosing more appropriate function names. In contrast, variable names seem to require a deeper analysis of the code's complexities, which might be a current limitation of the model.

Furthermore, the lower effectiveness scores, compared to the meaningfulness scores for both functions and variables, indicate a challenge for the model in establishing logical connections between different parts of the code. This highlights a potential area for improvement in the model's ability to understand and reflect code elements' specific roles and interactions.

For a detailed breakdown of the evaluation rubrics, please refer to the project documentation \cite{pordanesh2023}.

\subsubsection{Scenario 3}
In this experiment phase, as previously outlined in our experimental design, we subjected both the original and decompiled code to a series of eight Yes/No questions. These questions were sent to the LLM on each code sample, and the responses derived from the original code were used as benchmarks for comparison against the responses obtained from the decompiled codes.

\begin{table}[h]
\centering
\caption{Analysis of Question-Based Responses}
\label{table:question-based}
\begin{tabular}{cccc}
\toprule
Question number & Number of wrong answer & \% wrong answer \\
\midrule
1 & 0/15 & 0.0\% \\
2 & 4/15 & 26.7\% \\
3 & 3/15 & 20.0\% \\
4 & 5/15 & 33.3\% \\
5 & 5/15 & 33.3\% \\
6 & 3/15 & 20.0\% \\
7 & 7/15 & 46.7\% \\
8 & 4/15 & 26.7\% \\
TOTAL & 31/120 & 25.62\% \\
\bottomrule
\end{tabular}
\end{table}

\begin{table}[H]
\centering
\caption{Analysis of Application-Based Responses}
\label{table:application-based}
\begin{tabular}{cccc}
\toprule
Application id & \# Wrong Answers & \% Wrong Answers \\
\midrule
1 & 2 & 25.0\% \\
2 & 2 & 25.0\% \\
3 & 2 & 25.0\% \\
4 & 0 & 0.0\% \\
5 & 2 & 25.0\% \\
6 & 5 & 62.5\% \\
7 & 2 & 25.0\% \\
8 & 4 & 50.0\% \\
9 & 0 & 0.0\% \\
10 & 3 & 37.5\% \\
11 & 0 & 0.0\% \\
12 & 0 & 0.0\% \\
13 & 4 & 50.0\% \\
14 & 5 & 62.5\% \\
15 & 1 & 12.5\% \\
\bottomrule
\end{tabular}
\end{table}

Two key tables of results support our analysis. Table \ref{table:question-based} details the frequency of incorrect responses for each question across the 15 code samples. This provides insights into the consistency and reliability of the LLM's analysis capabilities. Table \ref{table:application-based} focuses on the number of incorrect responses per individual code sample. This aspect of the analysis potentially reflects each code sample's varying complexity and clarity.

Our primary focus will be on the first table, as it offers a more controlled view of the LLM's response patterns to our specific questions. The second table's insights show how code complexity impacts the LLM's performance, which gives us less flexibility for modification, as any code can be passed to the model by the user.

\textbf{Question-Based Analysis}
\begin{itemize}
    \item \textbf{Function Identification (Q1):} GPT-4 performed flawlessly in identifying the main functions in the code across all malware samples. This indicates a strong capability to recognize structural elements of code.
    \item \textbf{Network-Related Program (Q2) \& Scalability (Q3):} Errors in these areas were relatively low (26.7\% and 20.0\%, respectively), suggesting some capability in understanding network-related aspects and scalability but with room for improvement.
    \item \textbf{Suspicious API Calls (Q4) \& Network Activity (Q5):} Higher error rates of 33.3\% in both questions highlight a challenge in accurately identifying potentially malicious API usage and unauthorized network activities.
    \item \textbf{Encryption Routines (Q6):} With a 20.0\% error rate, GPT-4 can detect unrelated encryption routines, although it is not entirely foolproof.
    \item \textbf{Stealth Techniques (Q7):} The highest error rate (46.7\%) was observed here. This suggests that GPT-4 struggles more significantly recognizing cybersecurity manipulation code implementation.
    \item \textbf{Persistence Mechanisms (Q8):} An error rate of 26.7\% indicates a moderate challenge in identifying code ensuring auto-start or background activity.
\end{itemize}
\textbf{Application-Based Analysis}
\begin{itemize}
    \item \textbf{Consistent Performance Across Some Applications:} Some applications (4, 9, 11, 12) showed no errors, highlighting instances where GPT-4 could accurately respond to all questions.
    \item \textbf{Variable Performance:} Other applications (6, 13, 14) had high error percentages (above 50\%), showing significant challenges in those specific cases. This variance suggests that GPT -4's performance heavily depends on the specific nature of the code it analyzes.
    \item \textbf{Overall Error Rate:} An overall error rate of 25.62\% across all questions and applications suggests that while GPT-4 is generally reliable, there are significant areas where its analysis does not align with the benchmarked correct responses.
\end{itemize}

GPT-4 shows strong capabilities in understanding essential structural elements of code, such as function identification. The model struggles with some delicate aspects of code analysis, particularly in detecting stealth techniques and certain malicious activities. Also, the second table of results shows variable performance across different applications, suggesting that the complexity of the code may influence GPT -4's effectiveness.

\subsubsection{Scenario 4}

In this section, we delve into the model's capability to comprehensively explain decompiled code, which is crucial for initial reverse engineering analysis. We assessed the model's performance across four essential perspectives (Table \ref{table:code-explanation}), using the explanations from the original code as benchmarks. Each aspect was scored on a scale of 0 to 5 to quantify the model's performance in presenting primary information related to reverse engineering tasks.

\begin{table}[H]
\centering
\caption{Performance of GPT-4 in code explanation including 4 different perspectives}
\label{table:code-explanation}
\begin{tabular}{l c}
\toprule
\textbf{Aspect} & \textbf{Average Score (out of 5)} \\
\midrule
1. Functionality Overview & 2.4 \\
2. Key Observations & 1.73 \\
3. Security Analysis & 2.13 \\
4. Clarity and Structure & 2.33 \\
\bottomrule
\end{tabular}
\end{table}

In Functionality Overview, the GPT-4 displayed a moderate understanding of the code's primary functions, with an average score of 2.4. This indicates a reasonable capability to grasp basic functionalities but reveals inconsistencies, especially in complex codes.

Moving forward on Key Observations, scoring 1.73 on average, the model struggled in the key observations category. This reflects challenges in providing in-depth technical insights, which are crucial in reverse engineering.

With an average score of 2.13 in security analysis, the LLM showed a mixed ability to identify security risks. It could detect obvious vulnerabilities in some instances but lacked consistency across various code samples.

Also, in Clarity and Structure, the model's average score of 2.33 in clarity and structure suggests that while explanations were generally logical, they sometimes lacked organization, especially in complex scenarios.

Overall, GPT -4's performance varied across different aspects of code explanation. It was relatively better at explaining basic functionalities but less effective at detailed technical and security analyses. This suggests that while GPT-4 can assist in code explanation, it currently requires human expertise for more complex, nuanced tasks, particularly in security-critical scenarios.

\subsubsection{Scenario 5}

In this scenario, we presented the GPT-4 with original and decompiled code, asking it to answer seven targeted short-answer questions (Table \ref{table:short-answer}). These questions were designed to probe the model's depth of understanding and ability to extract and articulate key aspects of the code. We used the answers derived from the original code as a benchmark to evaluate the GPT -4's responses to the decompiled code.

\begin{table}[H]
\centering
\caption{Performance on answering 7 reverse engineering-related short answer questions}
\label{table:short-answer}
\begin{tabular}{l c}
\toprule
\textbf{Aspect} & \textbf{Average Score (out of 5)} \\
\midrule
1. Primary Functionality & 3.04 \\
2. Key Functions Description & 3.43 \\
3. Role of Selected Variable & 2.21 \\
4. Error Handling Mechanism & 2.93 \\
5. Flow of Execution & 2.71 \\
6. External Libraries/Dependencies & 2.68 \\
7. Evident Security Concerns & 2.64 \\
\bottomrule
\end{tabular}
\end{table}

Here is a breakdown of the results analysis of answering 7 short answer questions, which were scored manually based on a rubric:

\begin{enumerate}
\item \textbf{Primary Functionality (Avg. Score: 3.04):} We are supposed to have a score higher than other sections here, as in previous scenarios, we witnessed that the model's ability to provide a general overview of the code application is better than analyzing details.
\item \textbf{Key Functions Description (Avg. Score: 3.43):} In this section, we have got the best score, which is mainly about how the model can identify the main functions of the code. This is another task that focuses on a more comprehensive view of the application and fewer details.
\item \textbf{Role of Selected Variable (Avg. Score: 2.21):} The lower scores in this category indicate challenges in understanding the nuanced roles of individual variables within the code. However, variables in the original and decompiled codes vary, and we will talk more about it in the limitation and discussion sections.
\item \textbf{Error Handling Mechanism (Avg. Score: 2.93):} The LLM demonstrated an average understanding of error handling mechanisms in original and decompiled codes.
\item \textbf{Flow of Execution (Avg. Score: 2.71):} The model's ability to trace and explain the code's execution flow was average, highlighting potential difficulties in understanding complex control structures. The flow of execution can be a good parameter for comparison, as the final purpose of the application in both the original and decompiled codes is pretty similar.
\item \textbf{External Libraries/Dependencies (Avg. Score: 2.68):} The scores here reflect a fair ability to identify external dependencies, suggesting an understanding of the code's broader context. We are supposed to get a better score here as the main libraries for both code versions need to see some similarities, so we identified this task more accessible than others.
\item \textbf{Evident Security Concerns (Avg. Score: 2.64):} The model was moderately effective in identifying security issues, indicating a need for enhanced capabilities in security analysis. However, the question is ambiguous by itself, which makes answering this question hard for LLM. It can be interpreted as overall security issues, which the code may conduct as an application after compilation, or just security flaws on the code itself, making it vulnerable to manipulations.
\end{enumerate}

The results from this scenario suggest that while the LLM is average at extracting and explaining high-level functionalities and key functions, it struggles with more nuanced aspects, such as variable roles and detailed execution flows. Its performance in recognizing external libraries and pinpointing security concerns was adequate but indicated potential areas for further development.

We can discuss the comparison and questionnaire themselves, which may have caused this scenario to be a bit nonstandard for giving us a sound vision of GPT -4's performance on such a detailed analysis, which will be discussed later.

\section{Discussion and Limitations}

\subsection{Experiment Results}
In this discussion, we reflect on the experimental methods, evaluation techniques, and the implications of the results for understanding the performance of large language models (LLMs), in this case GPT-4, in reverse engineering tasks. However, the following sections will discuss doubts about our assumptions about experimental methods and evaluation techniques. 

The first phase's results were not within a reasonable range for meaningful discussion, highlighting evaluation limitations we will explore further. However, this phase was instrumental in shaping our approach to subsequent phases, particularly in data collection and experiment implementation. We discovered the complexity of data preparation, focusing on easily understandable samples with in-code comments and pre-written explanations to manage complexity effectively.

In the second phase, our findings offer more substantial insights. The first scenario demonstrated the GPT -4's proficiency in renaming variables and functions in a contextually relevant manner. While the scores for meaningfulness were higher, the effectiveness ratings also indicated a degree of reliability. This suggests potential areas for improvement in LLMs, particularly in establishing a reasonable channel of relations throughout the code components before choosing a new variable or function's name.

Scenario 3 revealed limitations in the GPT -4's accuracy, especially in addressing network-related and security-specific queries. This reflects the GPT -4's training on general code datasets and underscores the need for specialized, reverse engineering-focused datasets to enhance understanding in these areas.

The analysis of the fourth scenario indicated a shortfall in providing general explanations from reverse engineering viewpoints. Despite achieving less than 50\% accuracy across all subsections, there were instances of better performance in sections requiring a broad overview of the code. This suggests a potential reliance on the LLMs for identifying general application purposes, though it struggled with a more detailed network (cybersecurity perspectives) and structural questions.

Finally, scenario 5, which delved into detailed explanations, mirrored the earlier patterns: stronger performance in broader functionality queries and a weaker grasp of intricate code relationships. Interestingly, responses involving "functions" proceed better than those about "variables," and the model was more adept in identifying error handling mechanisms, a common feature in network applications, rather than manipulation mechanism detection tasks, which are more related to reverse engineering. This again points to the need for more reverse engineering-specific training to enhance the LLM's ability to distinguish and interpret complex coding patterns.

\subsection{Methods of Experiment }
In analyzing our research methods and their applicability to evaluating LLM performance in reverse engineering tasks, we confront various challenges and insights that will shape future research directions. Any discussion here will help us to find a better systematic way of experimenting in the next stage of our research journey.

During the first phase, the limitations of the BLEU score \cite{bleu} as an evaluation tool became evident. Originally designed for translation tasks, the BLEU score's effectiveness diminishes in assessing technical, computer science-related content. Its algorithmic focus on word-for-word comparisons fails to capture the nuanced understanding necessary for interpreting complex technical explanations. Thus, while BLEU scores may have a place in this research domain, their utility is restricted to scenarios with highly specific and limited explanatory expectations, which was not the case in our study.

The second phase's initial scenarios (1 and 2) showcased a more reliable evaluation method. These scenarios benefitted from direct human judgment with scoring based on a well-defined rubric. This approach underscored the value of human evaluators in contexts where nuanced understanding and contextual interpretation are crucial.

However, scenarios 3, 4, and 5 faced challenges due to their foundational assumption of the LLM's accuracy in interpreting original code. This assumption overlooks the inherent differences between human-written code and its decompiled equivalent. The distinct structural and methodological variations between these code forms, although having the same result in execution after compilation, mean that equivalent questions can yield differing yet valid responses for both codes, which makes these responses less comparable. Furthermore, our question selection could have been more effective. A focus on broader, general code aspects would likely yield more meaningful comparisons between original and decompiled code. This is especially relevant when our questions do not fully align with this standard.

Data collection presented another significant challenge, particularly in the second phase, where our focus on C code malware with existing comments and documentation restricted our data sources. Relying primarily on GitHub \cite{github} repositories limited the diversity and reliability of our dataset, thus impacting the generalizability of our findings. Additionally, the inconsistent nature of the dataset contributed to an uneven experimental landscape.

In addition, budget constraints related to the use of the GPT-4 API further limited our research scope, restricting the number of iterations we could afford.

All in all, our experience highlights the superiority of manual assessments in this context. The absence of specialized, automated evaluation tools for such complex tasks requires a more indicative approach. Ideally, a comparative analysis should involve two versions of decompiled code: one raw and one reviewed by expert reverse engineers. This benchmarked code, complete with expert annotations, could serve as a reference point for evaluating LLM responses. Developing a tailored questionnaire derived from the annotations on the benchmarked decompiled code would enable a more precise assessment of the LLM's capabilities in this delicate field. This data set would solve both valid benchmark and dataset constraints.

\section{Conclusion and Future Works}
The findings from our comprehensive investigation into the use of LLMs, particularly GPT-4 \cite{gpt4}, in Binary Reverse Engineering tasks reveal both the potential and the limitations of these advanced models. The initial phase of the study highlighted the challenges in automated evaluation methods, prompting a shift towards manual evaluation in subsequent phases. This change in approach facilitated a more in-depth understanding of LLMs' capabilities in code explanation and interpretation.

In the second phase, GPT-4 demonstrated a noteworthy ability to provide contextually relevant names for functions and variables and a general understanding of code functionalities. However, the model's performance varied significantly when delving into more complex aspects such as detailed technical insights, security analysis, and understanding intricate code relationships. Notably, the model showed stronger capabilities in explaining broader functionalities and identifying key functions compared to its performance in nuanced aspects like variable roles and specific execution flows.

The research also brought to light the limitations of existing evaluation methodologies in this domain. The BLEU score \cite{bleu}, while useful in certain contexts, proved inadequate for assessing complex technical explanations. Furthermore, the reliance on human-written code as a benchmark for comparison with decompiled code introduced inherent challenges due to structural differences between these code forms.

In future research, our focus will shift toward developing more sophisticated evaluation tools specifically designed for technical language analysis, particularly in the context of reverse engineering. Recognizing the need for broader and more varied datasets, efforts will be directed toward expanding the range of code types and sources. This will help create a comprehensive platform to test and understand the full potential of large language models (LLMs). Additionally, we plan to undertake comparative studies using expert-reviewed decompiled code as benchmarks. Addressing the constraints of resources, particularly in terms of budget and access to LLMs, will also direct us to use other open-source models for our examination.

\ifCLASSOPTIONcaptionsoff
  \newpage
\fi

\printbibliography

\end{document}